\documentclass[12pt,preprint]{aastex}
\usepackage{emulateapj5}
\usepackage{apjfonts}
\usepackage{epsfig}

\submitted{ApJL, accepted}

\def\ltsima{$\; \buildrel < \over \sim \;$}
\def\simlt{\lower.5ex\hbox{\ltsima}} % < over ~
\def\gtsima{$\; \buildrel > \over \sim \;$}
\def\simgt{\lower.5ex\hbox{\gtsima}} % > over ~

\shorttitle{Optical/UV Emission in 3C~279 Jet} 
\shortauthors{Cheung}

\begin{document}

\title{Detection of Optical Synchrotron Emission from the Radio Jet of 
3C~279}

\author{C.~C. Cheung}
\affil{Department of Physics, MS~057, Brandeis University, Waltham, MA
02454}
\email{ccheung@brandeis.edu}

\begin{abstract}

We report the detection of optical and ultraviolet emission from the
kiloparsec scale jet of the well-known quasar 3C~279. A bright knot,
discovered in archival V and U band {\it Hubble Space Telescope} Faint
Object Camera images, is coincident with a peak in the radio jet
$\sim$0.6$\arcsec$ from the nucleus. The detection was also confirmed in
Wide Field Planetary Camera-2 images. Archival Very Large Array and MERLIN
radio data are also analyzed which help to show that the high-energy
optical/UV continuum, and spectrum, are consistent with a synchrotron
origin from the same population of relativistic electrons responsible for
the radio emission.

\end{abstract}

\keywords{Galaxies: active --- galaxies: jets --- quasars: general ---
quasars: individual (3C~279) --- radio continuum: galaxies}

\section{Introduction}

The quasar 3C~279 (z=0.536) is one of the best studied active galaxies in
the sky. It is a bright EGRET $\gamma$-ray source \citep{hart01}, and
shows rapid variability across the entire electromagnetic spectrum
\citep[e.g.,][]{weh98}. One of the first early triumphs of the technique
of Very Long Baseline Interferometry (VLBI), was the discovery of apparent
superluminal motion in the parsec-scale jet of this quasar \citep[][and
references therein]{cot79}.  Subsequent superluminal ejections have been
continually monitored for more than thirty years
\citep[e.g.,][]{unw89,weh01}. The VLBI jet's polarization structure
\citep{lep95} has been imaged at frequencies as high as 86 GHz
\citep{att01}. Circular polarization has been detected in the parsec-scale
jet, possibly implying an electron-positron composition
\citep{war98,hom99}.

The radio jet has long been known to extend out to kiloparsec scales for
$\sim$5$\arcsec$ to the south-southwest \citep{dep83,aku94}.  An extended
feature to the northwest can be attributed to a radio lobe fueled by the
Doppler dimmed counter-jet, but is equally likely to be due to the main
jet because of its extreme viewing angle to our line of sight
\citep[Figure~\ref{fig-1};][]{dep83}. The recent discovery of an X-ray
counterpart to the 5$\arcsec$ long radio jet in a deep {\it Chandra}
exposure prompted an analysis of archival {\it Hubble Space
Telescope}\footnote{Based on observations made with the NASA/ESA Hubble
Space Telescope, obtained from the data archive at the Space Telescope
Science Institute. STScI is operated by the Association of Universities
for Research in Astronomy, Inc. under NASA contract NAS 5-26555.} ({\it
HST}) images in search of optical emission in the jet.  The {\it Chandra}
results will be presented elsewhere \citep{mar02b}. In this {\it Letter},
we report the detection of optical and ultraviolet (UV) emission
associated with the brightest region in the kiloparsec scale radio jet. We
adopt a Friedmann cosmology with $H_{0}$ = 75 km s$^{-1}$ Mpc$^{-1}$ and
$q_{0}$ = 0 throughout this {\it Letter}, so at the distance of the
quasar, 1 arcsecond equals 5.58 kpc.

\section{Description of Archival Observations}

We obtained post-COSTAR Faint Object Camera (FOC) f/96 \citep{jed94}
images of 3C~279 from the {\it HST} archive with `on-the-fly' pipeline
calibration applied. Single high quality exposures with the F253M and
F550M filters (each $\sim$1~ksec) were taken consecutively during November
1995 as part of a program (ID 6057) to study the inner structure of the
quasar \citep{hoo00} with the Fine Guidance Sensors (FGS) -- see Table~1.  
These filters approximate the standard U and V bands, respectively. To our
knowledge, only the FGS data have been published (E. Schreier, personal
communication).

\placetable{table-1}

\begin{figure*}
\figurenum{1}
\begin{center}
\epsfig{file=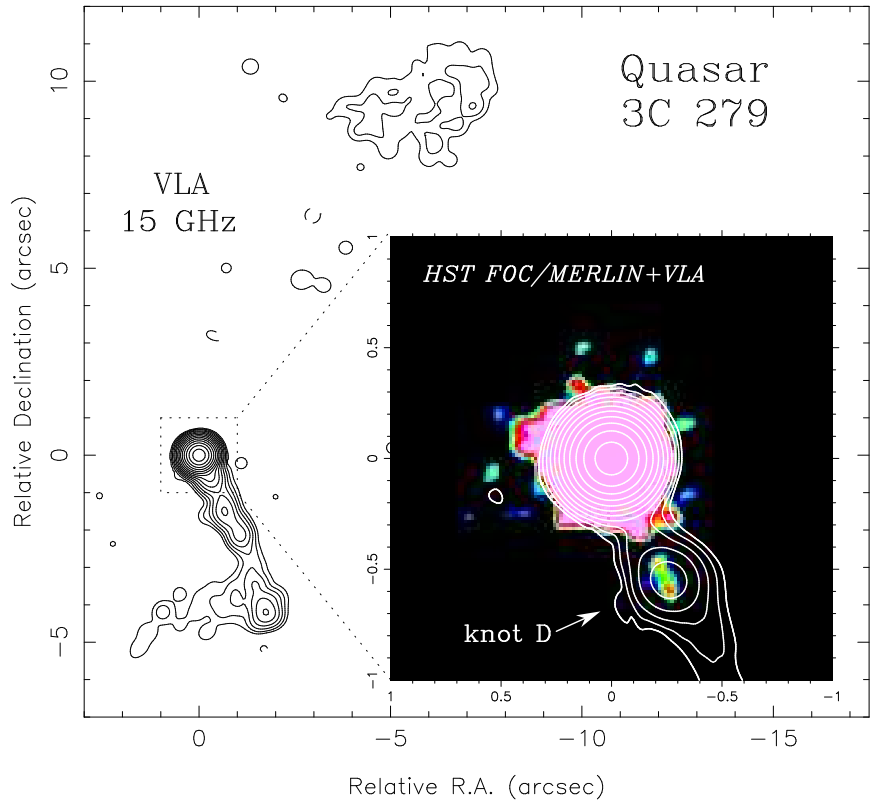,width=4.5in,angle=0}
\end{center}
\figcaption[f1.eps]{\label{fig-1} 
VLA 15 GHz image (resolution 0.4$\arcsec$) showing the large-scale radio
jet of 3C~279 at a position angle of $\sim205^{\circ}$, and a radio lobe
extending to the NW $\sim$10$\arcsec$ away. The inset is a zoomed in view
of the inner 2$\arcsec$$\times$2$\arcsec$ field around the nucleus. The
{\it HST} FOC V-band image (colorscale) was smoothed to a resultant
resolution of about 0.08$\arcsec$ and is shown with a high resolution
(0.2$\arcsec$) MERLIN+VLA 5 GHz
image overlaid (see Figure~\ref{fig-2} for the full radio image and
description). The images were aligned on the peaks of the nucleus. There
is a clear correspondence of an optical feature with a peak in the radio
jet emission at 0.6$\arcsec$ \citep[knot D -- naming convention
from][]{dep83}. The 15 GHz fluxes are plotted logarithmically in steps of
factor 2 beginning at 1.05 mJy/beam.}
\end{figure*}

A visual inspection of the FOC images shows a noticeable linear extension
centered $\sim$0.6$\arcsec$ from the quasar nucleus, at a position angle
(in the sky) of $\sim205^{\circ}$ in both the U and V band exposures. This
feature matches the position of the bright knot `D' in the radio jet
\citep[nomenclature of][]{dep83}.  An overlay of the optical image with a
high resolution radio image (Figure~\ref{fig-1}) show that the optical
feature is clearly coincident with this region of the radio jet.

\begin{figure*}
\figurenum{2}
\begin{center}
\epsfig{file=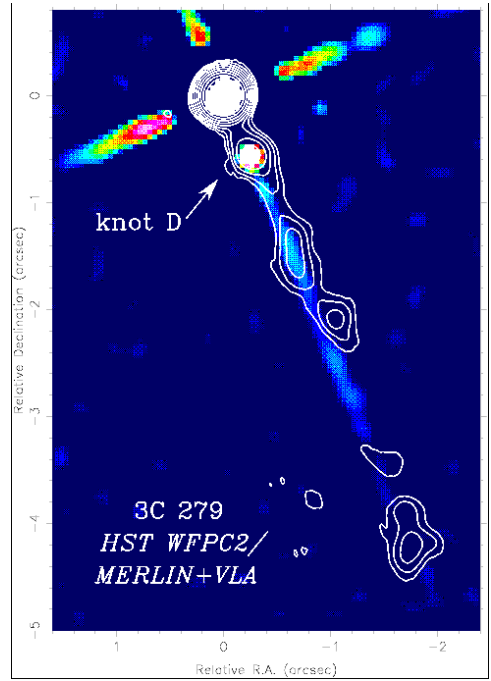,width=2.5in,angle=0}
\end{center}
\figcaption[f2.eps]{\label{fig-2}
{\it HST} WFPC2 F702W (R-band) image (greyscale plotted logarithmically)  
with MERLIN+VLA 5 GHz contour image restored with 0.2$\arcsec$ 
beam overlaid (see also Figure~\ref{fig-1}).
Optical emission from knot D appears prominently above a diffraction spike
after subtraction of a smoothed PSF from the image. The three other
diffraction trails extending radially from the nucleus are also apparent.
We were unable to identify any optical emission associated with the rest
of the radio jet due partly to the unfortunate positioning of the
diffraction spike at the position angle of the jet. The radio contours 
are plotted logarithmically in steps of factor 2 starting at 6 mJy/beam.}
\end{figure*}

We were able to confirm this detection in two short (300 sec) R-band WFPC2
F702W exposures taken more than a year apart as part of the {\it HST} 3CR
snapshot survey \citep{leh99}. Both images unfortunately suffered from
saturated nuclei, and even worse, a diffraction spike lies at the position
angle of the radio jet for almost its entire 5$\arcsec$ extent. However,
subtracting a smoothed version of the PSF from both the individual
exposures, and from the combined image, revealed a large excess above the
diffraction spike at the position of knot D in each case
(Figure~\ref{fig-2}). Its count rate (measured from the unsharp masked
image) is about three times greater than that measured anywhere else on
that or the other three diffraction spikes.  We use these data only for
confirming the optical identification of knot D, as any flux measurement
from these images will be hopelessly contaminated.

Two more {\it HST} images were available in the archive at the time of our
analysis from the same FOC/FGS program described above \citep{hoo00}. The
observing sequence from the original program was repeated $\sim$1.5 years
later (on 30 May 1997), but utilized the F2ND neutral density filters in
parallel which resulted in a factor of about six drop in sensitivity. Our
extraction procedure was performed on the images from the later epoch and
showed flux levels at the predicted position of the 0.6$\arcsec$ knot to
be consistent with the background (see \S{\ref{s:results}}). Fluxes for
the quasar nucleus could still be extracted from these images yielding 1.1
and 3.5 mJy at U and V bands, which are consistent with contemporaneous
ground-based measurements \citep{weh98}. We similarly extracted nuclear
fluxes for the 1995 epoch and obtained 0.79 mJy (U) and 2.8 mJy (V),
again, close to ground-based results by \citet{hart01}.

%This dominated over the $\sim2.8\%$ per year drop in sensitivity of the 
%FOC over its lifetime (\citet{jed97}; R. Jedrzejewski, personal 
%communication)

Radio data was obtained from the NRAO\footnote{The National Radio
Astronomy Observatory is a facility of the National Science Foundation
operated under cooperative agreement by Associated Universities, Inc.}
Very Large Array \citep{tho80} archive and the MERLIN\footnote{MERLIN is a
UK National Facility operated by the University of Manchester at Jodrell
Bank Observatory on behalf of PPARC.} archive (see Table~1 for a summary).
The VLA data (5 and 15 GHz) was calibrated in AIPS \citep{gre88} utilizing
standard procedures and self-calibrated in the Caltech DIFMAP package
\citep{she94}. We selected the data carefully to give us matched
resolution images ($\sim$0.4$\arcsec$) at the two frequencies. Fluxes were
set using the VLA 1999.2 coefficients for 3C~286. The MERLIN data (5 GHz)
was pipeline processed at Jodrell Bank on the \citet{baa77} scale, and
then combined at Brandeis with a day-long VLA observation recently
obtained when the full array was used as a single element for a VSOP
space-VLBI experiment.  We accounted for variability in the source between
the two epochs of observation by subtracting 9.75 Jy from the VLA nucleus
in the $(u,v)$ plane. With baselines up to 217 km, this data produced our
highest resolution radio image (Figures~\ref{fig-1} \&~\ref{fig-2}), which
most closely matches that achieved by the {\it HST} data
($\sim$0.05-0.1$\arcsec$).  \citet{pil87} presented an earlier image of
3C~279 using MERLIN only data. To our knowledge, only the WFPC2 data
\citep{leh99}, but none of the other multi-wavelength observations
presented in this work were previously published.

\section{Analysis and Results}
\label{s:results}

The jet emission in the two 1995 FOC exposures appear prominently above
the background and wings of the point spread function (PSF). Nevertheless,
in order to verify that the extended emission detected was not an artifact
of the PSF wings due to the bright nucleus, we subtracted scaled PSFs from
each 3C~279 image using images of stars from FOC calibration program ID
6196 (the FOC is insensitive to cosmic rays and a careful inspection of
the images did not show any unusual defects near the knot position). The
emission from radio knot D clearly remained in the resultant images (not
shown).  We did not detect any optical counterpart to the rest of the
arcsecond-scale jet\footnote{Up to $\sim$3$\arcsec$ where the field of
view of the FOC images was cut off, and for its entire length in the WFPC2
images.}, where the radio emission is weaker and the radio-to-optical
spectrum (based on previous ground based optical imaging) is expected to
be steeper: $\alpha_{ro}>1.1$ \citep{fra91}, using the sign convention
$F_{\nu}\propto\nu^{-\alpha}$. Analysis of the radio data and optical
limits for the rest of the jet is deferred to a future paper
\citep{mar02b}.

We extracted counts from the original (unrotated) geometrically corrected
and flat-fielded images using square apertures with 21 pixel
($\sim$0.3$\arcsec$) sides, centered on the position of the knot (we chose
not to extract fluxes from the PSF subtracted images since additional
artifacts were introduced from this process). This aperture was judged by
eye to contain all the flux from the jet knot, and is situated far enough
away from the bright quasar to avoid large contaminations.  The background
and contribution from the PSF was determined using the same aperture to
measure the average count-rates at the three 0.6$\arcsec$ distant
azimuthal positions in $90^{\circ}$ intervals around the core.  The count
rates were converted to flux densities using the inverse sensitivity
measurement given by the \verb+PHOTFLAM+ keyword in the image headers,
yielding 6.1 and 1.8 $\mu$Jy at the V and U bands, respectively. We
experimented with various circular apertures and found measurement
differences on the order of $\sim$20-25$\%$ (higher for the UV data) with
that derived above.  Therefore, we conservatively estimate this to be the
error in our {\it HST} flux measurements. These fluxes are consistent with
(1$\sigma$) upper limits of about 10 $\mu$Jy (V) and 6 $\mu$Jy (U)  
measured from the Poisson noise in the second epoch FOC data.

\begin{figure*}
\figurenum{3}
\begin{center}
\epsfig{file=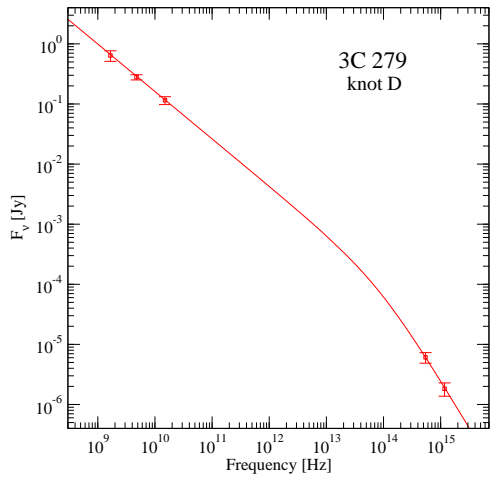,width=2.5in,angle=0}
\end{center}
\figcaption[f3.eps]{\label{fig-3}
The integrated radio-to-UV spectrum of the $\sim$0.6$\arcsec$ distant
region `D' in the kiloparsec scale jet of 3C~279. We fit a smoothly
connected double power law spectrum to the data with a break frequency at
$\sim9\times10^{13}$ Hz. See text for fluxes and associated errors.}
\end{figure*}

We used DIFMAP's \verb+MODELFIT+ program to fit elliptical gaussian models
to the radio visibility data in the inner region of the jet. The position
and size (including ellipticity) derived in this manner for knot D was
consistent between the frequencies and gave fluxes of 279 mJy (4.76 GHz)
and 115 mJy (15 GHz) which we estimate are accurate to better than 10 and
15$\%$, respectively. The size of the emitting region is best constrained
by the MERLIN+VLA image which gives a measured gaussian with dimensions
$0.28\arcsec\times0.06\arcsec$, elongated in the jet direction. We
estimated a 1.67 GHz flux (650 mJy $\pm20\%$) from a published MERLIN map
\citep{aku94} of comparable resolution to our new 5 and 15 GHz radio data,
giving us a three point {\it radio} spectral index ($\alpha_{r}$=0.79)  
for region D which agrees with the value of 0.75$\pm$0.1 measured by
\citet{dep83}. The optical and UV fluxes lie below the power-law
extrapolation from the radio measurements, indicating steepening at higher
frequencies where we measure an optical/UV spectral index of 1.6 with a
2-$\sigma$ confidence of 0.3. The radio-to-UV spectrum is shown in
Figure~\ref{fig-3}.

\section{Physical Conditions in the Optical/UV Emitting Region}

The radio emission from region D shows a high level of fractional
polarization (20$\%$) at 5 GHz, but does not appear to be strongly shocked
as the magnetic field direction is predominantly longitudinal
\citep{dep83} indicating a shear dominated flow. Its integrated
radio-to-UV spectrum can be well explained by a steepening synchrotron
spectrum from a single population of relativistic electrons, as is the
usual interpretation for the growing number of already detected optical/UV
arcsecond-scale jets \citep{spa94,sca02}. We can extrapolate the radio
spectrum to the optical/UV by fitting a break at $\sim9\times10^{13}$ Hz
to a smoothly connected double power-law spectrum (Figure~\ref{fig-3}), so
the synchrotron luminosity appears to peak in the near-infrared. If we can
extend the observed spectrum to even higher energies, it would predict a 1
keV flux of roughly 0.4~nJy if the X-ray output from the 0.6$\arcsec$
feature is dominated by synchrotron losses.  This should be easily
detectable in the {\it Chandra} exposure obtained by \citet{mar02b},
although it may be difficult to resolve with the detector's limited
resolution of $0.5\arcsec$/pixel.  The measured steepening in the spectral
index ($\Delta\alpha\sim0.8$) corresponds to a change in the electron
distribution index $\Delta$$p$ (see below) of 1.6. Similar changes have
been found for other well-studied high-energy (optical and X-ray)
synchrotron emitting regions in the jets of, for instance, M87
\citep{mar02a} and PKS~0521--365 \citep{bir02}.

We applied a homogeneous synchrotron model \citep[see e.g.,][]{blu70} to
calculate an equipartition magnetic field,
$B_{eq}\simeq1\times10^{-3}~\times~[\eta^{0.26}~\gamma_{min}^{-0.15}~f^{-0.26}~\delta^{-1}]~G$.
The factor $\eta$ is defined by the relation: $u_{B}=\eta~u_{e}$, where
$u_{B}$ and $u_{e}$ are the energy densities in the magnetic field and
electrons, respectively (e.g., if we have equipartition, $\eta=1$ for an
electron-positron plasma).  We assumed that the electrons have a power law
energy distribution $N(\gamma)\propto\gamma^{-p}$ (with minimum and
maximum energy cutoffs, $\gamma_{min}<\gamma<\gamma_{max}$, and
$p=2\alpha+1$), and that the radiating particles and magnetic field fill a
fraction, $f$, of a spherical volume whose diameter is $4\times10^{21}$ cm
(applying an appropriate conversion from our measured elliptical gaussian
fit to the size; \citet{mar83}). The Doppler beaming factor is defined as
$\delta \equiv [\Gamma (1-\beta\cos\theta)]^{-1}$, where $\Gamma$ is the
bulk Lorentz factor of the knot, $\beta$ is the velocity of the jet in
units of c, and $\theta$ is the angle the jet makes to our line of sight.
This calculation is only weakly dependent on $\gamma_{max}$ which we
arbitrarily set to extend into infinity.  In the extreme case that the
errors in the radio fluxes were even larger than our conservative
estimates, it may be possible to obtain a single power law index
($\alpha\sim$1) fit to the overall radio-to-UV spectrum. Our calculation
of $B_{eq}$ in this case maintains relatively unaffected by this steeper
spectral index. The derived magnetic field implies that the UV emitting
electrons have
$\gamma\simeq8\times10^{5}~\times~[\eta^{-0.13}~\gamma_{min}^{0.08}~f^{0.13}]$
with a radiative lifetime (in their rest frame) of about
$20~\times~[\eta^{-0.39}~\gamma_{min}^{0.22}~f^{0.39}~\delta^{2}$] yrs.

It is useful to see how the synchrotron cooling time of the UV emitting
electrons in knot D compares with the light travel distance from the
nucleus. Firstly, the angle between the jet and our line of sight is
restricted to be $\leq8^{\circ}$ by the observed VLBI apparent
superluminal motion of 14c \citep{cot79}. We note that this measurement is
the fastest (and earliest) motion recorded in this quasar so far, and was
observed when the VLBI jet was best aligned ($218^{\circ}\pm2^{\circ}$)
with the position angle of the VLA scale jet ($\sim205^{\circ}$). This
gives us a deprojected distance from the nucleus to knot D of $\simgt$25
kpc. Now, if the Doppler factor this far down the jet is as large as that
observed on parsec-scales \citep[$\delta=20$;][]{hom00}, and making the
simple assumptions, $\eta=1$, $\gamma_{min}=10$, and $f=1$, we can infer a
rest-frame $B_{eq}\sim4\times10^{-5}$ G -- comparable to those derived for
other known optical emitting synchrotron features \citep[see
e.g.,][]{sca02}. It follows that the {\it observed} (accounting for time
dilation between the rest frame and observer frame) radiative lifetime is
$\sim3\times10^{5}$ yrs which corresponds to a travel distance of $\sim$90
kpc and exceeds the light travel time from the nucleus by a factor of a
few.  This makes it just plausible that {\it in-situ} reacceleration
\citep[e.g.,][]{mei96} is not strictly required if the Doppler factor at
the site of knot D is really as high as that measured on parsec scales,
but that remains unknown.

\acknowledgements

The author is grateful to John Wardle for a careful reading of the
manuscript and continued support of his research, to Herman Marshall for
communications which inspired the {\it HST} archive search and support of
this work, and to Dave Roberts, Dan Homan, Niruj Mohan, Fabrizio
Tavecchio, Robert Jedrzejewski, and Andy McDonald for invaluable
discussions. Radio astronomy at Brandeis University is supported by the
National Science Foundation through grants AST 98-02708 and AST 99-00723.

%%%%%%%%%%%%%%%%%%%%%%
%%                  %%  
%%     Tables+Figs  %%
%%                  %%
%%%%%%%%%%%%%%%%%%%%%%

%\include{table1}
%\include{table2}

\begin{deluxetable}{lccc}
\label{table-1}
\tablecolumns{4} 
\tablewidth{0pc} 
\tablecaption{Summary of Archival Observations\label{tbl-1}}
\tablehead{ 
\colhead{Instrument}            &\colhead{Date}
&\colhead{Frequency}            &\colhead{Exp Time}             
\\ 
\colhead{(1)}                   &\colhead{(2)}
&\colhead{(3)}                  &\colhead{(4)}                
}  
\startdata 
MERLIN        & 04 Jan 1993 &4.99 &$\sim$9 hrs   \\
VLA A-configuration& 10 Jan 2001 &4.87 &$\sim$7 hrs \\
VLA A-configuration& 12 Mar 1990 &4.76 &580 sec  \\
VLA B-configuration& 06 Jan 1992 &15.0 &2690 sec \\
{\it HST} WFPC2/F702W& 23 Feb 1994 &4.33$\times$10$^{5}$ & 300 sec \\
{\it HST} WFPC2/F702W& 20 May 1995 &4.33$\times$10$^{5}$ & 300 sec \\
{\it HST} FOC/F550M & 29 Nov 1995 &5.44$\times$10$^{5}$ &1115 sec\\
{\it HST} FOC/F253M & 29 Nov 1995 &1.17$\times$10$^{6}$ & 935 sec\\ 
\enddata 
\tablecomments{Column (1) indicates the telescope and 
instrument/configuration used.
The MERLIN observations utilized a six element array (no Lovell). 
The WFPC2 data was not used for photometric analysis.\\
Column (2) is the date of observation in UT. \\
Column (3) is the frequency in GHz. \\
Column (4) is the time on source with approximate times indicated for 
heavily edited data.
}
\end{deluxetable}


\begin{thebibliography}{}

\bibitem[Akujor et al.(1994)]{aku94} Akujor, C.~E., L{\"u}dke, E., Browne,
I.~W.~A., Leahy, J.~P., Garrington, S.~T., Jackson, N., \& Thomasson, P.\
1994, \aaps, 105, 247

\bibitem[Attridge(2001)]{att01} Attridge, J.~M.\ 2001, \apjl, 553, L31

\bibitem[Baars et al.(1977)]{baa77} Baars, J.~W.~M., Genzel, R.,
Pauliny-Toth, I.~I.~K., \& Witzel, A.\ 1977, \aap, 61, 99

\bibitem[Birkinshaw, Worrall, \& Hardcastle(2002)]{bir02} Birkinshaw, M.,
Worrall, D.~M.,~\& Hardcastle, M.~J.\ 2002, \mnras, 335, 142

\bibitem[Blumenthal \& Gould(1970)]{blu70} Blumenthal, G.~R.~\& Gould,
R.~J.\ 1970, Rev. Mod. Phys., 42, 237

%\bibitem[Celotti, Ghisellini, \& Chiaberge(2001)]{cel01} Celotti, A.,
%Ghisellini, G., \& Chiaberge, M.\ 2001, \mnras, 321, L1

\bibitem[Cotton et al.(1979)]{cot79} Cotton, W.~D.~et al.\ 1979, \apjl,
229, L115

\bibitem[de Pater \& Perley(1983)]{dep83} de Pater, I.~\& Perley, R.~A.\
1983, \apj, 273, 64

\bibitem[Fraix-Burnet et al.(1991)]{fra91} Fraix-Burnet, D., Golombek, D.,
Macchetto, F., Nieto, J.-L., Lelievre, G., Perryman, M.~A.~C., \& di
Serego Alighieri, S.\ 1991, \aj, 101, 88

\bibitem[Greisen(1988)]{gre88} Greisen, E.~W. 1988, AIPS Memo 61, National 
Radio Astronomy Observatory

%\bibitem[Harris \& Krawczynski(2002)]{har02} Harris, D.~E.~\& 
%Krawczynski, H.\ 2002, \apj, 565, 244

\bibitem[Hartman et al.(2001)]{hart01} Hartman, R.~C., Boettcher, M.,
Aldering, G. ~et al.\ 2001, \apj, 553, 683

\bibitem[Homan \& Wardle(1999)]{hom99} Homan, D.~C.~\& Wardle, J.~F.~C.\
1999, \aj, 118, 1942

\bibitem[Homan \& Wardle(2000)]{hom00} -----.\ 2000, \apj, 535, 575

\bibitem[Hook, Schreier, \& Miley(2000)]{hoo00} Hook, R.~N., Schreier,
E.~J., \& Miley, G.\ 2000, \apj, 536, 308

\bibitem[Jedrzejewski et al.(1994)]{jed94} Jedrzejewski, R.~I., Hartig,
G., Jakobsen, P., Crocker, J.~H., \& Ford, H.~C.\ 1994, \apjl, 435, L7

\bibitem[Lehnert et al.(1999)]{leh99} Lehnert, M.~D., Miley, G.~K.,
Sparks, W.~B., Baum, S.~A., Biretta, J., Golombek, D., de Koff, S.,
Macchetto, F.~D., \& McCarthy, P.~J.\ 1999, \apjs, 123, 351

\bibitem[Lepp{\"a}nen, Zensus, \& Diamond(1995)]{lep95} Lepp{\"a}nen,
K.~J., Zensus, J.~A., \& Diamond, P.~J.\ 1995, \aj, 110, 2479

\bibitem[Marscher(1983)]{mar83} Marscher, A.~P.\ 1983, \apj, 264, 296

\bibitem[Marshall et al.(2002a)]{mar02a} Marshall, H.~L., Miller, B.~P.,
Davis, D.~S., Perlman, E.~S., Wise, M., Canizares, C.~R., \& Harris,
D.~E.\ 2002a, \apj, 564, 683

\bibitem[Marshall et al.(2002b)]{mar02b} Marshall, H.~L. et al. 2002b, in preparation

\bibitem[Meisenheimer(1996)]{mei96} Meisenheimer, K.\ 1996, In "Jets from
Stars and Active Galactic Nuclei," W.~Kundt (ed.), Springer Lecture Notes
No. 471, 57

\bibitem[Pilbratt, Booth, \& Porcas(1987)]{pil87} Pilbratt, G., Booth,
R.~S., \& Porcas, R.~W.\ 1987, \aap, 173, 12

%\bibitem[Piner et al.(2000)]{pin00} Piner, B.~G., Edwards, P.~G.,
%Wehrle, A.~E., Hirabayashi, H., Lovell, J.~E.~J., \& Unwin, S.~C.\ 2000,
%\apj, 537, 91

%\bibitem[Sambruna et al.(2002)]{sam02} Sambruna, R.~M., Maraschi, L.,
%Tavecchio, F., Urry, C.~M., Cheung, C.~C., Chartas, G., Scarpa, R., \&
%Gambill, J.~K.\ 2002, \apj, 571, 206

%\bibitem[Scarpa et al.(1999)]{sca99} Scarpa, R., Urry, C.~M., Falomo, 
%R., \& Treves, A.\ 1999, \apj, 526, 643

%\bibitem[Schwartz et al.(2000)]{sch00} Schwartz, D.~A., Marshall, H.~L.,
%Lovell, J.~E.~J.~et al.\ 2000, \apjl, 540, L69

\bibitem[Scarpa \& Urry(2002)]{sca02} Scarpa, R.~\& Urry, C.~M.\ 2002,
New Astronomy Review, 46, 405

\bibitem[Shepherd, Pearson, \& Taylor(1994)]{she94} Shepherd, M.~C.,
Pearson, T.~J., \& Taylor, G.~B. 1994, BAAS, 26, 987

\bibitem[Sparks, Biretta, \& Macchetto(1994)]{spa94} Sparks, W.~B.,
Biretta, J.~A., \& Macchetto, F.\ 1994, \apjs, 90, 909

%\bibitem[Tavecchio et al.(2000)]{tav00} Tavecchio, F., Maraschi, L.,
%Sambruna, R.~M., \& Urry, C.~M.\ 2000, \apjl, 544, L23

\bibitem[Thompson et al.(1980)]{tho80} Thompson, A.~R., Clark, B.~G.,
Wade, C.~M., \& Napier, P.~J.\ 1980, \apjs, 44, 151

\bibitem[Unwin et al.(1989)]{unw89} Unwin, S.~C., Cohen, M.~H., Hodges,
M.~W., Zensus, J.~A., \& Biretta, J.~A.\ 1989, \apj, 340, 117

%\bibitem[Wardle \& Aaron(1997)]{war97} Wardle, J.~F.~C.~\& Aaron, S.~E.\
%1997, \mnras, 286, 425

\bibitem[Wardle et al.(1998)]{war98} Wardle, J.~F.~C., Homan, D.~C., Ojha,
R., \& Roberts, D.~H.\ 1998, \nat, 395, 457

\bibitem[Wehrle et al.(1998)]{weh98} Wehrle, A.~E., Pian, E., Urry,
C.~M.~et al.\ 1998, \apj, 497, 178

\bibitem[Wehrle et al.(2001)]{weh01} Wehrle, A.~E., Piner, B.~G., Unwin,
S.~C., Zook, A.~C., Xu, W., Marscher, A.~P., Ter{\"a}sranta, H., \&
Valtaoja, E.\ 2001, \apjs, 133, 297

\end{thebibliography}
\end{document}